\documentstyle[prl,aps,multicol,epsf]{revtex}

\title{\Large {\bf Electron localization in one dimension
obtained  from  combined exact diagonalization 
- {\sl ab initio} approach}}
\author{ Jozef  Spa\l{}ek and Adam Rycerz }
\address{Marian Smoluchowski Institute of Physics, Jagiellonian University, ulica Reymonta 4, \\
30-059 Krak\'{o}w, Poland\\}
\date{June 4, 2001}

\newcommand{\half}%
{\frac{\scriptstyle 1}{\scriptstyle 2}}

\begin{document}
\maketitle
\begin{abstract}
Exact ground-state properties 
are presented by combining the diagonalization in the Fock 
space (and taking {\it all} hopping integrals and {\it all} 
two-site interactions)
with the {\it ab initio} optimization of the Wannier
functions. 
Electrons are essentially localized for the interatomic
distance $R \sim 2.0$ \AA\  for $s$-like states,
when the quasiparticle mass is divergent. 
The momentum distribution {\it dispersion}  
is proposed to define the localization {\it order parameter}.
Dimerization and zero-point energies are also discussed. 
The method
provides convergent results for $N\geq 8$ atoms.\\
\vspace{0in} 
\end{abstract}
% ********************
\begin{multicols}{2}
\narrowtext

One dimensional systems range from organic
metals \cite{jero} to quantum rings and wires \cite{jaca},  
and to nanotubes \cite{mint}. 
In their description the role
of the long-range Coulomb interaction is crucial because
of reduced dimensionality, for which the charge screening becomes
less effective \cite{ovch}. 
The existing exact solutions of the parametrized
models with inclusion of intersite interactions \cite{ovch,stra} 
prove 
the existence of the metal-insulator transition for the half-filled-band
case, in contradistinction to the corresponding Hubbard-model
solution \cite{lieb}, for which the system is insulating even for 
an arbitrarily
small Coulomb repulsion. 
The existence of such metal-insulator the transition has
been also discussed \cite{daul} within the density-matrix renormalization
group (DMRG) 
method when the second-neighbor hopping is included. 
A separate
question concerns the appearance of the Tomonaga-Luttinger
behavior \cite{emer} in the metallic state, for which some evidence has
been gathered \cite{dard}. 
In brief, the search for a proper description
of those systems as quantum liquids (and their instabilities) is one
of the basic problems in the physics of low-dimensional systems.

In the above theoretical analysis [1-8] the solutions
have been disscussed as a function of the microscopic
parameters, which are not easy to measure. Therefore, one assumes
that they should be determined first from a~separate single-particle approach.
In following this route one must avoid counting
twice the interaction, as discussed carefully in the papers 
implementing the LDA+U \cite{anis} and SIC \cite{temm} methods.
We have proposed \cite{spapo} a new method in which the 
single-particle (Wannier) wave functions are allowed to 
relax in the correlated state and thus are determined by optimizing 
the exact ground state energy obtained from the diagonalization in the
Fock space.
In other words, we treat properly the interactions first due
to their strongly nonperturbative nature and then 
readjust the single-particle functions $\{w_i({\bf r})\}$
by setting Euler equation for them, which plays the role 
of the renormalized wave equation for a particle
in the correlated ground state. 
In such  approach the problem of double counting 
the interaction does not arise {\it at all}. 
Additionally, we include 
{\it all} the hopping integrals and {\it all} two-site interactions
to make the solution more complete. 
The method is executable on
a desktop server for the number 
of atoms $N\leq 12$. 
What is probably the most remarkable formal
feature of these calculations is that 
the electronic correlations make the wave function more tightly 
bound to the atoms and thus limiting the interaction range. 
In effect, the results are converging very fast for $N\geq 8$ atoms 
meaning that the {\it optimized} Wannier functions and the interaction 
parameters are relevant at most up to the {\it third} neighbors. 
The  
results are detailed below and to the best of our knowledge 
they represent the first analysis of the strong correlation 
effects as a function of the lattice parameter 
within the exact acount of both the interaction and the wave function
{\it without limiting
the range of either the hopping processes or the two-site interactions}.
The obtained results show that the method is particularly useful for accurate
studies of electronic correlations in quantum dots and rings.

\begin{figure}
\centering\epsfxsize=8cm\epsffile{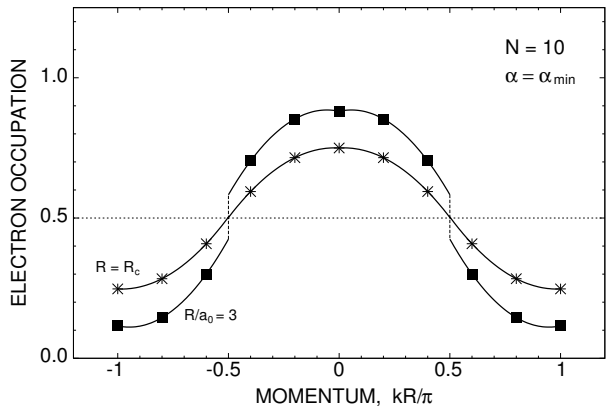} \par
\caption{
Momentum distribution $n_{k\sigma}$ for electrons in linear ring 
of $N=10$ atoms; the interatomic distance $R$ is specified in 
units of Bohr radius $a_0$. 
The continuous line represents the parabolic interpolation,
which is of the same type for both $k>k_F$ and $k<k_F$.}
\end{figure}

The nature of the electron momentum distribution \cite{carmelo}
$n_{k\sigma}\,\equiv\,<a_{k \sigma}^{\dagger}\,a_{k \sigma}>$ 
is shown in Fig.~1  for $N=10$ atoms (we use the periodic
boundary conditions). For the interatomic distance $R=3a_0$, where
$a_0$ is the $1s$ Bohr radius (setting the length scale), this is essentially 
the Fermi-Dirac 
function with  a tail extending to the Brillouin zone boundary. 
The smearing out of the distribution with increasing 
$R$ suggests that the electronic states transform 
from itinerant to localized states, 
as exemplified for $R=R_c\approx 3.929$.
The continuous line represents the parabolic fit of the same type 
for both $k\leq k_F$ and $k\geq k_F$. 
A natural question to be dealt with is whether the Fermi ridge 
disappearance at $R=R_c$ is reflected in any singularity in other properties. 

\begin{figure}
\centering\epsfxsize=8cm\epsffile{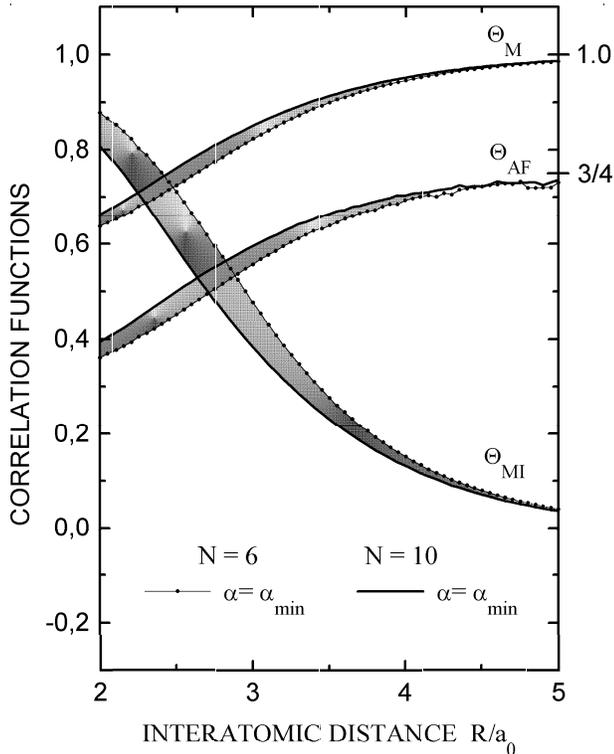} \par
\caption{
Correlations function {\it versus}  distance $R$,
depicting 
the crossover from itinerant to localized state (see main text), for
$N=6$ to $10$ atoms. The shaded areas are drawn to illustrate the convergence
of the results  in the large $R$ limit.}
\end{figure}

To distinguish quantitatively
between those states we plot
in Fig.~2, the following basic quantities (as a function of $R$) 
defined in both itinerant and localized states: 
(i) the site spin magnitude 
$\theta_M\equiv (4/3)\left<{\bf S}^2_i\right>$, where 
${\bf S}_i\,=\,(S_{i}^{+},\,S_{i}^{-},\,S_{i}^{z})\,
=\,(a_{i \uparrow}^{\dagger}\,a_{i \downarrow},\,
a_{i \downarrow}^{\dagger}\,a_{i \uparrow},\, (n_{i \uparrow}\,-\,
n_{i \downarrow})/\,2)$ 
is the electron spin 
on site $i$, (ii) the spin correlation function 
$\theta_{AF}=-\left<{\bf S}_i\cdot{\bf S}_{i+1}\right>$,  
and (iii) $\theta_{MI}=4{\sigma}^2\left\{n_{k\sigma}\right\}$,
where ${\sigma}^2\left\{n_{k\sigma}\right\}$ is 
the dispersion of the statistical distribution defined as:
\begin{equation}
  {\sigma}^2\left\{n_{k\sigma}\right\}=
  \frac{1}{2N}\sum_{k\sigma}n_{k\sigma}^2-
  \left(\frac{1}{2N}\sum_{k\sigma}n_{k\sigma}\right)^2.
\end{equation}
The averages are for the ground state, which is determined via an exact
diagonalization in the Fock space.
The quantity ($\theta_M\,=\,1\,-\,2\,<n_{i \uparrow}\,
n_{i \downarrow}>$) 
takes the value ($1/2$) in the ideal
gas limit 
and approaches unity in the atomic limit, where we have a Pauli spin
on each atom. 
$\Theta_{AF}$ approaches the value ($3/4$) for 
the singlet configuration of atomic spins, whereas
$\sigma^2\left\{n_{k\sigma}\right\}$
acquires the value $1$ in the gas limit
($n_{k \sigma}\,=\,\Theta\,(\mu\,-\,\epsilon_{k})$)
and vanishes for an even momentum distribution 
($n_{k \sigma}\,=\,1/2$), when  
the particle position is sharply defined on atom,
i.e. for the 
localized electron states characterized by atomic states. 
Thus, the quantity $\theta_{MI}$ plays the role
of {\it the order parameter} for this {\it crossover behavior},  
since it clearly 
distinguishes between the complementary momentum and position representations 
of the system quantum states. 
From Fig.~2 it 
follows that for $R/a_0=5$ all three parameters acquire 
(within  $5\%$) their asymptotic values for purely atomic states.
%This suggests that the localization is achieved in a continuous manner, as
%is also the case in any dynamical mean-field theory (DMFT)  
%\cite{bull}. Only slightly different values are obtained if we define
%a $1 \%$ criterion of localization (the exact number should be provided
%by specific experiment). 

\begin{figure}
\centering\epsfxsize=8cm\epsffile{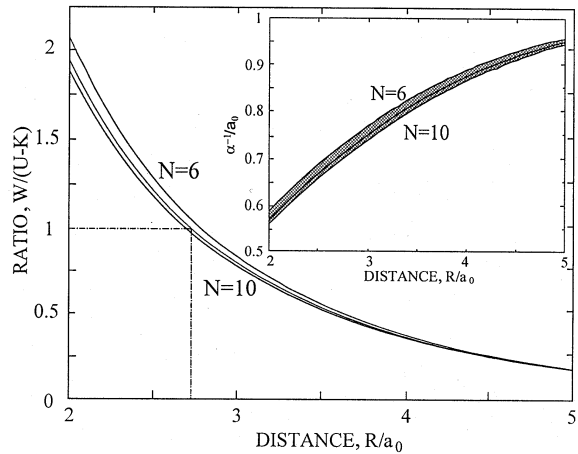} \par
\caption{
The bandwith-to-interaction ratio versus $R$; 
the Hubbard point for localization [15] is marked.
The inset provides the optimal size $\alpha^{-1}$ (in units of
$a_{0}$ for $s$ atomic orbitals
composing the {\it optimized} Wannier function.}
\end{figure}

The analysis of metal-insulator transition induced by electron-electron
interaction started with the works of Mott \cite{mott} and Hubbard \cite{hubb}.
The first of them introduced the critical carrier concentration for the
transition to take place in a discontinuous manner, whereas the second
introduced the critical bandwidth to interaction ratio (of the order of unity).
Our method of approach can be used to relate the above criteria, as we discuss
next. 
Namely, to relate our method  to the original ideas of 
Mott  and Hubbard, we have plotted in Fig.~3 
the ratio  of bandwidth $W \equiv 2 |\sum_{j(i)} t_{ij}|$, where $t_{ij}$ is 
the hopping integral between  the neighbors $\left<ij\right>$
(calculated  through  the relation  
$t_{ij}=\left<w_i\right|H_1\left|w_j\right>$ for
the basis $\left\{w_i\right\}$ optimized in the correlated state and for
full single-particle potential in $H_1$), to the
effective short range Coulomb interaction $U-K$, where 
$U=\left<w_iw_i\right|V_{12}\left|w_iw_i\right>$
and $K=\left<w_iw_{i+1}\right|V_{12}\left|w_iw_{i+1}\right>$. 
The value $W/\left(U-K\right)=1$
marked in this figure represents roughly the dividing 
line between metallic and Mott insulating states for
three-dimensional systems \cite{hubb}. 
This point, achieved
for $R\approx 2.7a_0$, does not reflect any characteristic  point 
for our system. 
Instead, the localization is practically achieved
for the distance about twice as large, as is the size of the atomic states
composing the Wannier function \cite{spapo} (characterized 
by the quantity ${\alpha}^{-1}$, see the inset), which nears   
its atomic value for the $s$-like state. 
Again, the results for $N=8$
and $N=10$ are very close to each other;
this is the reason why we have shadowed the areas 
between the corresponding curves in both Figs.~2 and 3.

On the basis of Figs.~2 and 3 we can
estimate the localization threshold for our system. 
The corresponding Mott criterion \cite{mott}, generalized to  
$d$ dimensions takes the form $n_c^{1/d}a_H\approx 0.2$, 
where $n_c$ is the carrier
concentration, and $a_H$ is the size of the states 
at the localization threshold. 
In our situation of  neutral chain:  $n_c=1/R$, and 
$a_H={\alpha}^{-1}$ so that for $R\approx 5a_0$ this criterion takes the 
form 
$\left(a_0/R\right)\left({\alpha}^{-1}/a_0\right)\approx 0.95/5\approx 0.2$,
a suprisingly close 
value to that of Mott (which reflects the long-range nature of the Coulomb
interaction). 
Thus, our results provide a support
for the Mott criterion rather than for the Hubbard one \cite{hubb}. In other
words, the {\it metallicity} extends well beyond the $W=U$ limit
and this must be caused by the inclusion of more distant hopping processes.
One may also say that the Mott-criterion universality 
originates from the long-range nature of the interaction, which
imitates the higher-lattice dimensionality.

%We have corroborated the {\it metallicity} criterion further by determining 
%directly the $R$ value at which the {\it Fermi ridge} in Fig.~1 disappears
%(and corresponds to the effective mass divergence in bulk systems
%\cite{brinkman} signalling the localization). Within the parabolic
%interpolation of the Fermi ridge $\Delta n_{k_{F}}(R)$
%we obtain $R\,\approx\,4\,a_{0}$. 
Hence, the conclusion \cite{lieb} about the universality
of the insulating state for the Hubbard chain does not 
extend to the $1d$ models with a realistic account of
the electronic structure.
This  conclusion is very important also because it removes
one of the main objections against using the itinerant 
(or even effective mass) states for 
quantum dots \cite{jaca}. 

\begin{figure}
\centering\epsfxsize=8cm\epsffile{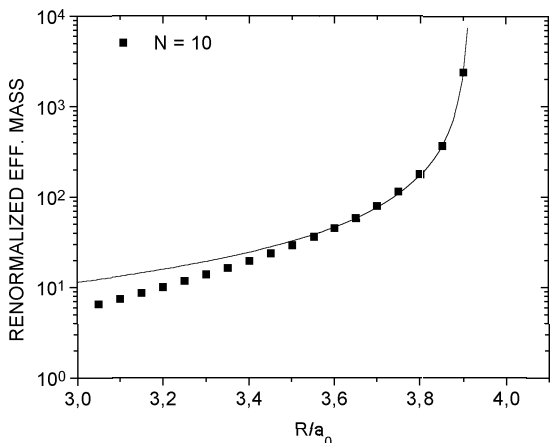} \par
\caption{
The critical behavior of the quasiparticle mass at the Fermi level:
The computed points are fitted with the curve discussed in the text.}
\end{figure}

We can determine directly the 
effective mass at the Fermi level. Namely, the calculated band mass 
$m_{F}$ is about $40\%$ enhanced near the localization threshold.
The quasiparticle mass $m^{\star}_{F}$ is found from the relation
$m^{\star}_{F} = (\Delta n_{k_{F}})^{-1} m_{F}$, where the first factor 
is the usual Fermi-liquid $Z^{-1}$ enhancement 
(the discontinuity at the Fermi level marked in Fig.1). 
The results are shown in Fig.~4.
A clear critical behavior is detected: $m^{\star}_{F} = A |R - R_{c}|
^{-\gamma}$, with $A = 10.2, R_{c} = 3.92$, and $\gamma = 4/3$. This quantum
critical behavior is obtained, since we emulate the discrete distribution
$n_{k}$ with a continuous parabolic interpolation when determining 
$\Delta n_{k_F}$. 
The localization threshold $R_{c}$ is about $10\%$
higher for $N=8$. It is tempting to speculate that with the increasing 
$N\rightarrow\infty$ the 
Mott and the Hubbard criteria for $R_{c}$ may coalesce. The masses near 
$R = R_{c}$ are huge and reach those known only for heavy-fermion compounds.

The one-dimensional systems are unstable with respect to the 
dimerization \cite{jero}. 
We have 
determined the ground state energy of such a state  
on the same level of precision as above and have 
additionally minimized it with respect to the amplitude of the lattice
distortion. 
In Fig.~5 (bottom) we have compared the energy contribution $\Delta E$
due to the dimerization with that due to the zero-point motion (top). 
Both the dimerization amplitude
and $\Delta E$ are strongly reduced for $R/a_0 >5$, 
a feature 
inevitably connected with the long-range nature of Coulomb
interactions, which drives the system towards spatially periodic
state (the Mott insulating state on the lattice replaces the Wigner-crystal 
state \cite{emer} for the electron gas). 
Therefore, our results and conclusions  above
should remain intact for longer chains. 
Additionally, the zero-point energy 
overcomes the dimerization energy for a light (hydrogen) 
chain and has been estimated in the following manner.

In the harmonic 
approximation, the phonons have energy
$\omega_k=2\left(C/M\right)^{1/2}\sin\left(\pi k/N\right)$, 
where $M$ is ionic mass, and the elastic constant is
calculated by a numerical differentiation 
$C=N^{-1}{\partial}^2E_G/\partial R^2$.
Formally, these formulae are valid only if the 
energy has an absolute minimum. 
Here we assume that the system is closed
in a box of length $NR$ and thus the global repulsive interaction
between the atoms (the situation with $\partial E_G/\partial R < 0$ for
given $R$) is balanced out  by the environment. 
Also, as the k=0 mode is a Goldstone mode, we
include only the modes with $k \neq 0$. The 
phononic contribution to the ground state energy (in the atomic
units) is then
\begin{equation}
  \Delta E_G^{\rm ph}=\left(\frac{2m}{M}\right)^{\half}
  \left(\frac{1}{N}\frac{{\partial}^2 E_G}{\partial R^2}\right)^{\half}
  \sum^{N-1}_{k=1}\sin\left(\frac{\pi k}{N}\right),
\end{equation}
where $m$ is the bare electron mass. On the 
basis of the relation for the $k$-th mode 
$\left(1/2\right)M\omega_k\left(\Delta R_k\right)^2=\hbar\omega_k/2$,
where $\left(\Delta R_k\right)^2$ is the mode contribution to the 
zero-point vibrations, we can easily estimate the total 
amplitude $\left(\Delta R\right)^2$, which in the 
atomic units has the form
\begin{equation}
  \left(\Delta R\right)^2= \frac{1}{N}\left(\frac{m}{2M}\right)^{\half}
  \left(\frac{1}{N}\frac{{\partial}^2E_G}{\partial R^2}\right)^{-\half}
  \sum_{k=1}^{N-1}\frac{1}{\sin\left(\pi k/N\right)}.
\end{equation}
In the $N\rightarrow\infty$ this result give 
$\left(\Delta R\right)^2\sim\log N$, providing the dynamical
lattice instability in one dimension. 
At the localization threshold (i.e. for $R/a_0\approx 5$) and for $N=8$, we
have that 
$\Delta R\approx 0.12$, a substantial.
By comparison, the 
shift due to the dimerization is  $\approx 0.06$ . 
Also, the {\it Peierls distorted state} extends 
to the localized state (up to $R\,\approx\,6.5\,a_{0}$). 

\begin{figure}
\centering\epsfxsize=8cm\epsffile{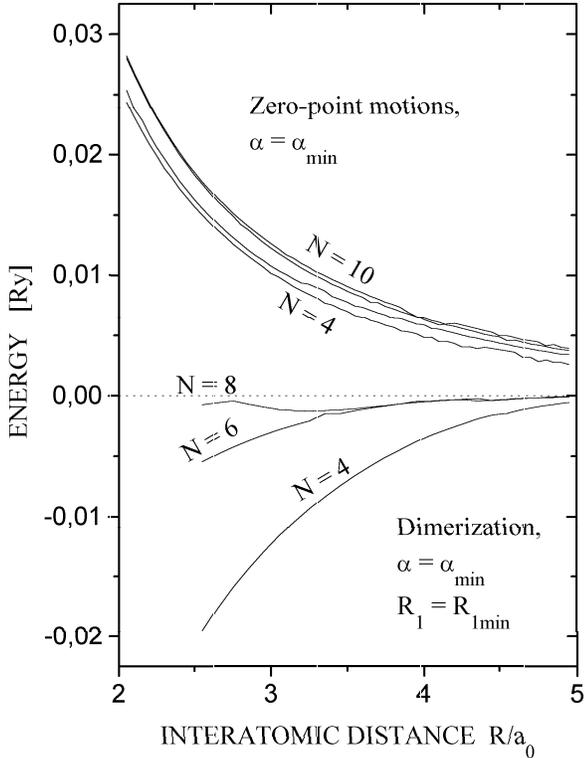} \par
\caption{
The zero-point  (top curves) and the dimerization
(bottom curves) energies for $N=4\div 10$ atoms {\it versus} 
$R$  ($\alpha_{min}^{-1}$ is  the optimal size
of the atomic orbitals).}
\end{figure}

In summary, we have determined the microscopic 
criteria for the {\it transition} from the itinerant to the localized states
in a one-dimensional system of a finite size
and have illustrated those findings on the 
example of a correlated quantum ring.
The new method of 
optimizing the single-particle wave functions in the correlated
state \cite{spapo} proves thus valuable
in the exact treatment of nanoscopic systems.
We have calculated
all the properties as a function of the lattice parameter.   
%The implementation of the method \cite{spapo} within DMFT 
%\cite{bull} would provide the mean-field solution of the Hubbard model as a
%function of the lattice parameter.
The work was supported by KBN Grant No. 2PO3B~092~18.

\end{multicols}

\end{document}